\documentclass{IEEEtran}

\usepackage{parskip}

\usepackage{graphicx}
\usepackage{dcolumn}
\usepackage{bm}
\usepackage{epsfig}
\usepackage{float}
\usepackage{amsmath}
\usepackage{amssymb}
\usepackage{tabularx}
\usepackage{tikz}
\usepackage{xcolor}
\usepackage{hyperref}
\usepackage{cleveref}
\usepackage{comment}
\usepackage{longtable}
\usepackage{url}
\usepackage{xspace}
\usepackage[all]{nowidow}
\usepackage[switch]{lineno}

\usepackage{cite}
\usepackage{amsmath,amssymb,amsfonts}
\usepackage{textcomp,nicefrac}

\def\BibTeX{{\rm B\kern-.05em{\sc i\kern-.025em b}\kern-.08em
T\kern-.1667em\lower.7ex\hbox{E}\kern-.125emX}}

\begin{document}

\title{Design and Beam Test Results for the 2D Projective sPHENIX Electromagnetic Calorimeter Prototype}

\author{
C.A.~Aidala, S.~Altaf, R.~Belmont, S.~Boose, D.~Cacace, M.~Connors, E.~Desmond, J.~Frantz, E.A.~Gamez, N.~Grau, J.S.~Haggerty, A.~Hodges, J.~Huang, Y.~Kim, M.D.~Lenz, W.~Lenz, N.A.~Lewis, E.J.~Mannel, J.D.~Osborn, D.V.~Perepelitsa, M.~Phipps, R.~Pisani, S.~Polizzo, A.~Pun, M.L.~Purschke, C.~Riedl, T.~Rinn, A.C.~Romero~Hernandez, M.~Sarsour, Z.~Shi, A.M.~Sickles, C.~Smith, S.~Stoll, X.~Sun, E.~Thorsland, F.~Vassalli, X.~Wang, C.L.~Woody

\thanks{
This research was carried out using resources from the Fermi National Accelerator Laboratory (Fermilab), a HEP user facility managed by the Fermi Research Alliance, LLC, for the U.S. Department of Energy, Office of Science, acting under Contract No. DE-AC02-07CH11359.\par
 
Please see Acknowledgements for author affiliations.
}
}

\maketitle
\begin{abstract}
sPHENIX is a new experiment under construction for the Relativistic Heavy Ion Collider at Brookhaven National Laboratory which will study the quark-gluon plasma to further the understanding of QCD matter and interactions. A prototype of the sPHENIX electromagnetic calorimeter (EMCal) was tested at the Fermilab Test Beam Facility in Spring 2018 as experiment T-1044. The EMCal prototype corresponds to a solid angle of $ \Delta \eta \times \Delta \phi = 0.2 \times 0.2$ centered at pseudo-rapidity $\eta = 1$. The prototype consists of scintillating fibers embedded in a mix of tungsten powder and epoxy. The fibers project back approximately to the center of the sPHENIX detector, giving 2D projectivity. The energy response of the EMCal prototype was studied as a function of position and input energy. The energy resolution of the EMCal prototype was obtained after applying a position dependent energy correction and a beam profile correction. Two separate position dependent corrections were considered. The EMCal energy resolution was found to be $\sigma(E)/\langle E\rangle = 3.5(0.1) \oplus 13.3(0.2)/\sqrt{E}$ based on the hodoscope position dependent correction, and $\sigma(E)/\langle E\rangle = 3.0(0.1) \oplus 15.4(0.3)/\sqrt{E}$ based on the cluster position dependent correction. These energy resolution results meet the requirements of the sPHENIX physics program.
\end{abstract}


\section{Introduction}
\label{sec:introduction}

sPHENIX is a new experiment \cite{Adare:2015kwa} under construction for the Relativistic Heavy Ion Collider at Brookhaven National Laboratory which will study the quark-gluon plasma (QGP) \cite{Aschenauer:2016our,Adcox:2004mh,Adams:2005dq,Back:2004je,Arsene:2004fa} to further the understanding of QCD matter and interactions. sPHENIX is designed to measure the QGP at a variety of length scales using various probes to provide insights into the microscopic properties of the QGP. One such probe is jets that arise from hard scattering interactions between two partons, with the energy loss of partons traversing the QGP being of particular interest. sPHENIX will allow for a detailed study of flavor dependent energy loss through a measurement of heavy flavor tagged jets, as well as open heavy flavor hadrons. Measurements of photon-tagged jets and jet substructure are also part of the sPHENIX physics program. sPHENIX will allow for measurements of jets with transverse momentum as low as 10 GeV/c, as well as provide measurements of both the hadronic and electromagnetic components of jets at RHIC. To accomplish these measurements, sPHENIX is designed with a tracking system, a calorimeter system with $2\pi$ azimuthal acceptance and pseudorapidity coverage of $|\eta| < 1.1$, and the former BaBar solenoid magnet \cite{OConnor:1998llb}. The calorimeter system consists of an electromagnetic calorimeter and a hadronic calorimeter. The use of the BaBar magnet imposed constraints on the sPHENIX detector design. In particular, the electromagnetic calorimeter was required to be compact enough to fit inside the magnet while allowing enough space for the tracking system and part of the hadronic calorimeter. The electromagnetic calorimeter was also designed to be compact in order to minimize the cost of the calorimeter system.

The sPHENIX electromagnetic calorimeter (EMCal) is a sampling calorimeter designed to measure the electrons, positrons, and photons in electromagnetic showers. The EMCal will also measure approximately one interaction length of hadronic showers. The EMCal has a coverage of $|\eta|<1.1$ and full azimuth. The EMCal is segmented into \emph{towers} of size $\Delta \eta \times \Delta \phi = 0.024 \times 0.024$, with an approximate volume of 2.5$\times$2.5$\times$14 cm$^3$, which sets the granularity of the calorimeter. The towers are defined within calorimeter \emph{blocks} that consist of scintillating fibers embedded in a mix of tungsten powder and epoxy. Each block corresponds to a 2$\times$2 array of towers. Each tower is equipped with a light guide coupled to silicon photomultipliers that collect the light from the fibers. The blocks are distributed in 64 sectors that describe an overall cylindrical geometry concentric with the beamline and centered at the interaction point of the particle collisions. Each side $0<|\eta|<1.1$ has 32 sectors distributed evenly in azimuth. Each sector has 24 rows of blocks extending along the beamline, and each row has 4 blocks along the $\phi$ direction. The blocks are tapered in both $\eta$ and $\phi$, resembling a truncated pyramid, and giving a 2D projective geometry. The blocks are further tilted such that the fibers do not project directly at the interaction point, minimizing channeling and improving energy resolution. More details about the sPHENIX detector and the EMCal can be found in reference \cite{TDR}.

\begin{figure*}[htb!]
\centering
\includegraphics[width=\textwidth]{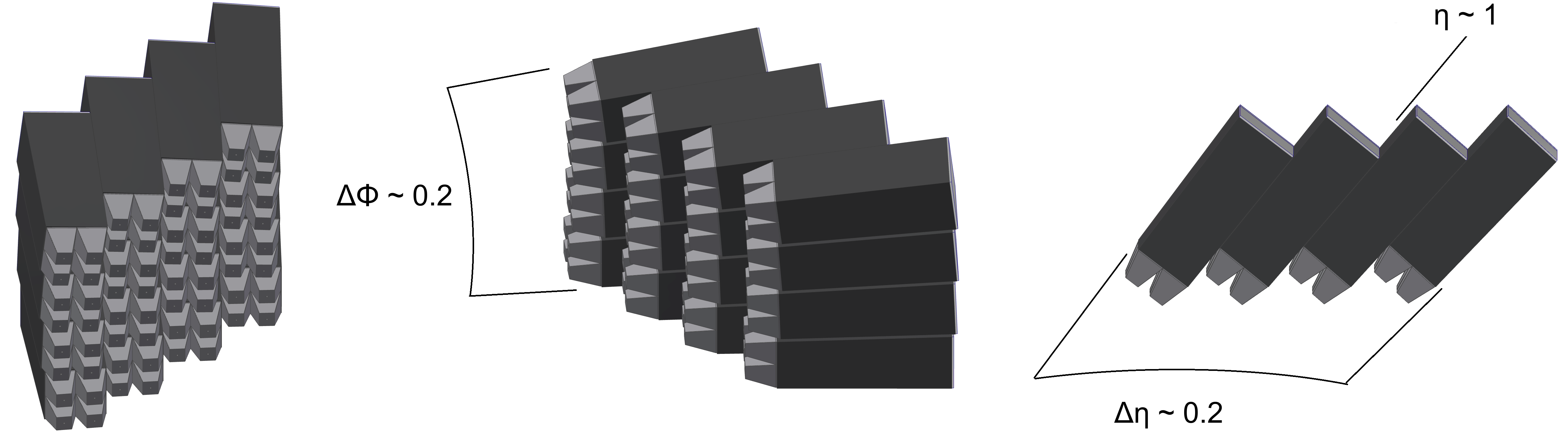}
\caption{EMCal prototype. The prototype consists of an array of 4$\times$4 blocks, covering a solid angle of $\Delta \eta \times \Delta \phi = 0.2 \times 0.2$ centered at $\eta = 1$. Each block (dark gray) corresponds to a 2$\times$2 array of towers defined by light guides (light gray).}\label{drawing_emcal}
\end{figure*}

The sPHENIX physics program requires an EMCal energy resolution equal or better than $16\%/\sqrt{E} \oplus 5\%$. This requirement is motivated by the measurement of the Upsilon states through the electronic decay channel $\Upsilon~\rightarrow~e^-e^+$. The electrons from these Upsilon decays are expected to produce EMCal electromagnetic showers with energies of approximately 4 to 10 GeV. In contrast, underlying event fluctuations in central Au+Au collisions would produce a comparable measurement of approximately 320 MeV \cite{TDR}. The energy resolution requirement was based on the maximum energy smearing that would allow discrimination of the Upsilon states against the average underlying event fluctuations.

A prototype of the EMCal was constructed in order to test its energy resolution. The prototype corresponded to an array of 8$\times$8 calorimeter towers, or 4$\times$4 blocks, centered at $\eta = 1$. The prototype covered a solid angle of $ \Delta \eta \times \Delta \phi = 0.2 \times 0.2$. Figure \ref{drawing_emcal} shows a schematic view of the EMCal prototype.

A previous prototype of the EMCal was tested in 2016 \cite{Aidala:2017rvg}. There are various differences between the 2016 prototype and the 2018 prototype discussed in this paper. The most notable difference is the projectivity of the EMCal blocks. The 2016 prototype was only 1D projective (in $\phi$), whereas the 2018 prototype is 2D projective (in $\eta$ and $\phi$). The 2D projectivity is a desirable feature because it improves energy measurements at higher pseudorapidity. For a 2D projective design, an electromagnetic shower at high pseudorapidity is contained within a smaller number of towers than for a 1D design, which results in a greater signal per tower and a better discrimination against underlying event fluctuations. Another difference between the prototypes is the pseudorapidity region that they covered. While both prototypes corresponded to a slice $\Delta \eta \times \Delta \phi = 0.2 \times 0.2$ of the EMCal, the 2016 prototype was centered at $\eta=0$ and the 2018 prototype was centered at $\eta=1$. The change in pseudorapidity was motivated by the fact that the 2D projectivity reduces to 1D towards $\eta=0$ because the sPHENIX detector is symmetric with respect to this plane. Other changes were also introduced in the 2018 prototype in order to optimize the EMCal design (details in reference \cite{TDR}), but the 2D projectivity and the high pseudorapidity are the main differences with respect to the previous prototype. The final EMCal design that will be implemented in sPHENIX will closely follow the design of the 2018 prototype.

\section{Prototype Electromagnetic Calorimeter}
\label{sec:emcal}

\subsection{EMCal Block Production}

The EMCal blocks were produced by embedding a matrix of scintillating fibers in a mix of epoxy and tungsten powder. The blocks are similar to the ``Spaghetti Calorimeter" design used in other experiments \cite{Tsai:2012cpa,Tsai:2015bna,Leverington:2008zz,Sedykh:2000ex,Armstrong:1998qs,Appuhn:1996na,Hertzog:1990md}. The scintillating fibers are as long as the block and are distributed uniformly across the block's cross section. There is a total of 2668 fibers per block. The towers within a block have an area of approximately $(1.1R_M)^2$, where $R_M \approx$ 2.3 cm is the Moli\`ere radius. The length of the towers varies with $\eta$ and it has an approximate value of 20$X_0$, where $X_0 \approx$ 7 mm is the radiation length. The block density is approximately 9.5 g/cm$^3$, with a sampling fraction of approximately 2.1$\%$.

\begin{table}[htb!]
\centering
\caption{EMCal block materials}
\begin{tabularx}{.48\textwidth}{llr}
\hline
\multicolumn{1}{c}{\bf Material} & \multicolumn{1}{c}{\bf Property} & \multicolumn{1}{c}{\bf Value} \\
\hline
\\
Scintillating fiber 	& Saint Gobain BCF-12 & \\
                        & diameter & 0.47 mm\\
                        & core material & polystyrene \\
                        & cladding material & acrylic \\
                 		& cladding & single\\
                 		& emission peak & 435 nm \\
                 		& decay time & 3.2 ns \\
                 		& attenuation length & $\ge$ 1.6 m \\
& & \\
Tungsten powder & THP Technon 100 mesh	&  \\
                & particle size	&  25-150 $\mu$m \\
                & bulk density (solid) & $\ge$ 18.50 g/cm$^3$ \\
                & tap density (powder) & $\ge$ 10.9 g/cm$^3$ \\
                & purity & $\ge$ 99\% W \\
                & impurities ($\le$ 1\%) & Fe, Ni, O$_2$, Co,\\
                & & Cr, Cu, Mo \\
& & \\
Epoxy                       & EPO-TEK 301 & \\
\hline
\end{tabularx}
\label{materials}
\end{table}

The materials used to produce the blocks are listed in Table \ref{materials} along with some of their properties. The blocks were produced at the University of Illinois at Urbana-Champaign following this procedure \cite{TDR}:

\begin{itemize}
    \item Scintillating fibers are dropped into mesh screens that hold the fibers in place.
    \item The fiber-screen assembly is put into a mold.
    \item Tungsten powder is poured into the mold. The mold is placed on a vibrating table to pack the powder.
    \item Epoxy is poured into the top of the filled mold, while a vacuum pump is used at the bottom to extract the air as well as pull the epoxy through the mold.
    \item The filled mold is left to dry until the mix is solid.
    \item The block is unmolded and machined to its final shape. A diamond tip is used to machine the readout ends of the block.
\end{itemize}

\begin{figure}[htb!]
\centering
\includegraphics[width=0.33\textwidth]{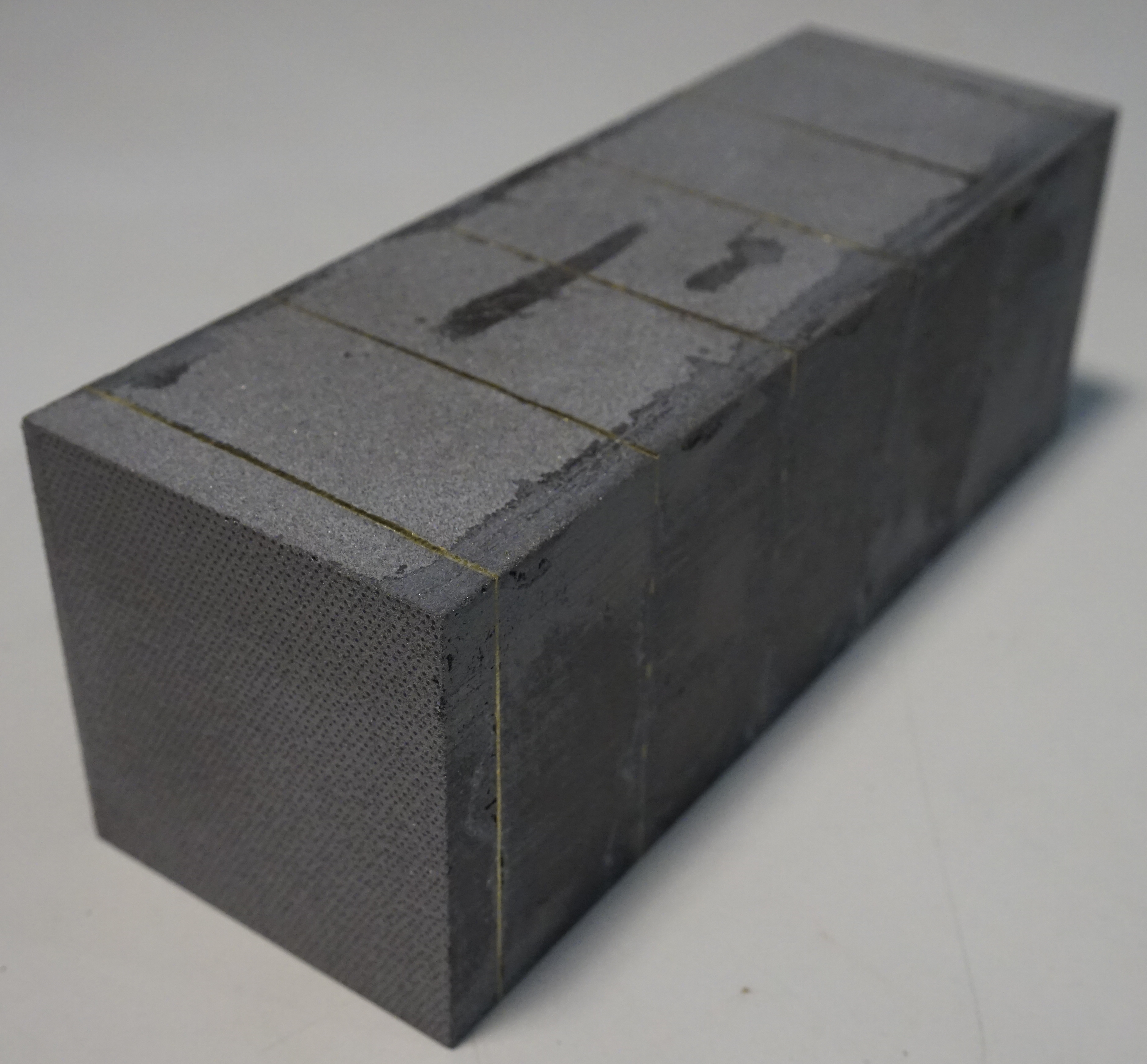}
\caption{EMCal block. The block consists of scintillating fibers embedded in a  mix of tungsten powder and epoxy. The blocks are tapered in two dimensions, giving a 2D projective geometry.} \label{block}
\end{figure}

A finished EMCal block can be seen in Figure \ref{block}. The quality assurance of the blocks included tests of density, light transmission and size. The blocks had a density ranging from 9.2 to 9.8 g/cm$^3$. All the blocks had more than 99$\%$ fibers that successfully transmitted light. The size of the blocks deviated from the nominal dimensions by less than 0.5 mm.

\subsection{Light Collection}
\begin{figure}[htb!]
\centering
\includegraphics[width=0.47\textwidth]{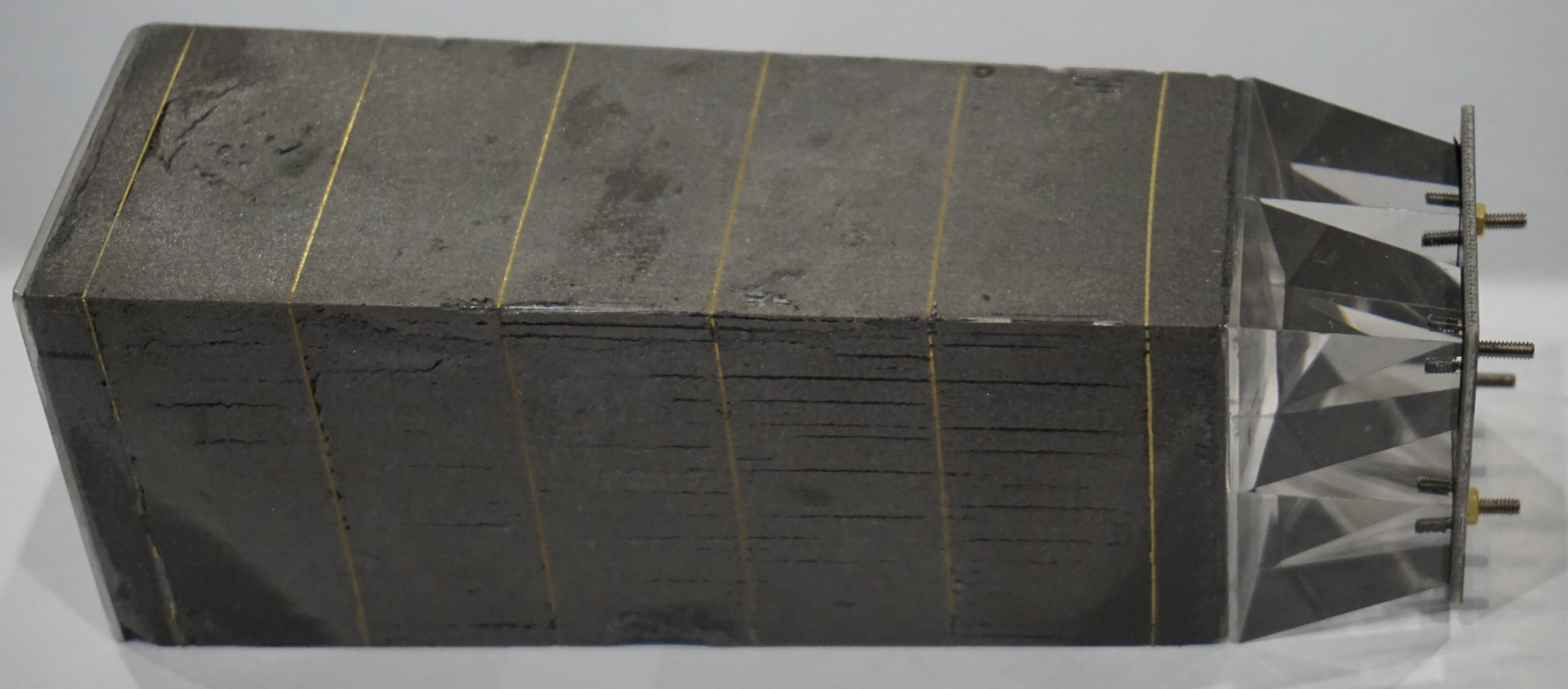}
\caption{EMCal block equipped with light guides and SiPMs.} \label{block_lightguides}
\end{figure}

The light from the scintillating fibers was collected at the tower's front end (closer to the interaction point). Light guides were epoxied to the front of the blocks, while aluminum reflectors were epoxied to the back. The light guides consisted of UV transmitting acrylic with a trapezoidal shape (see Figure \ref{block_lightguides}), custom made by NN, Inc.\footnote{NN, Inc., Precision Engineered Products Group, East Providence, RI~02914.} A silicone adhesive was used to couple each light guide to a 2$\times$2 array of silicon photomultipliers (SiPM). Each SiPM (Hamamatsu S12572-015P) had an active area of 3$\times$3 mm$^2$ containing 40K 15$\mu$m pixels, and had a photon detection efficiency of 25$\%$. The signals from each of the four SiPMs were summed to give a single output signal from each tower. More details about the electronics are given in Section \ref{sec:electronics}. Figure \ref{block_lightguides} shows an EMCal block equipped with light guides and SiPMs.

\subsection{Assembly}
\begin{figure}[htb!]
\centering
\includegraphics[width=0.47\textwidth]{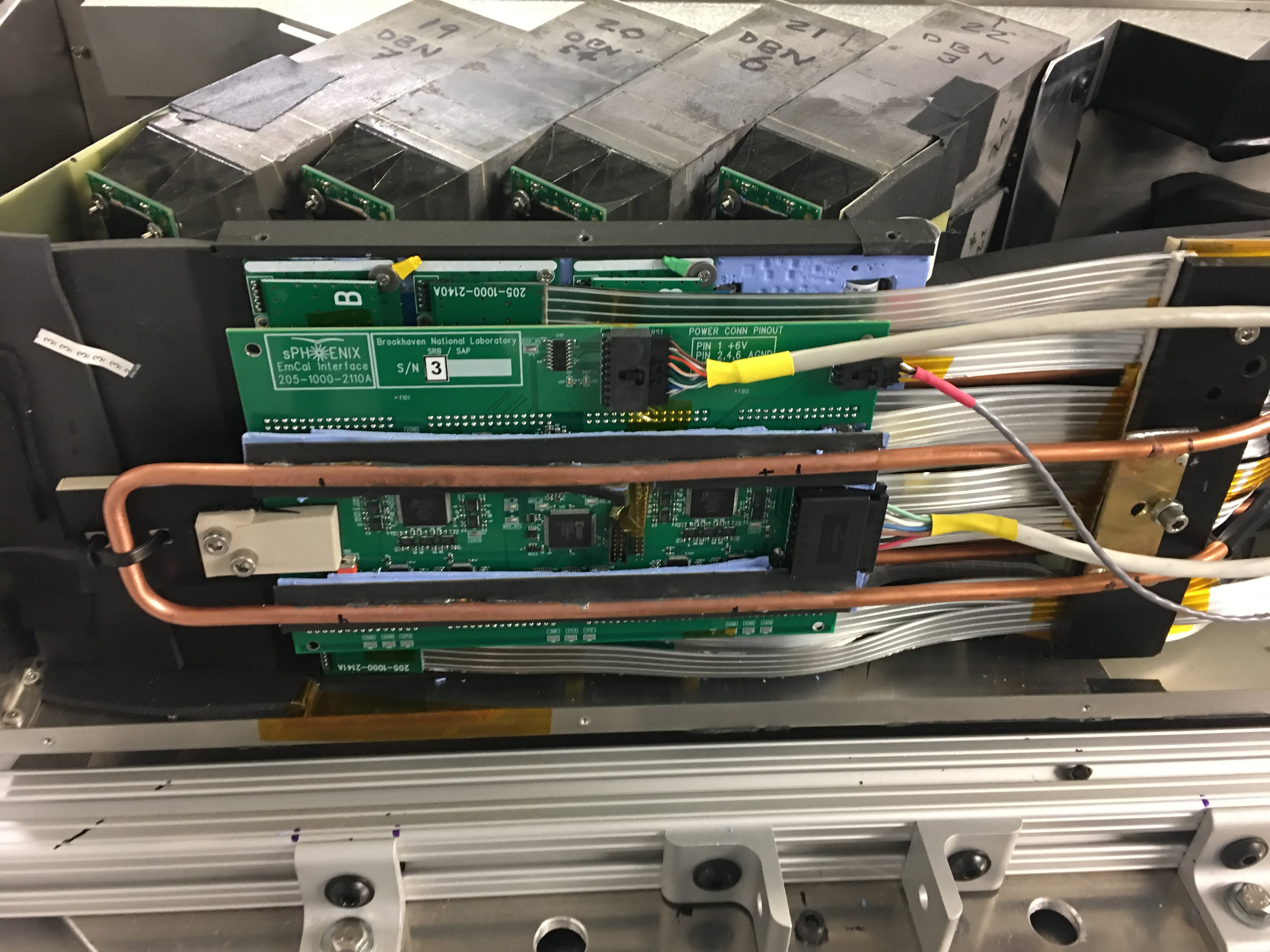}
\caption{EMCal prototype showing the EMCal blocks, light guides, SiPMs, electronics and part of the cooling system.} \label{setup_emcal}
\end{figure}

Once the EMCal blocks were equipped with light guides and SiPMs, they were stacked and epoxied together in their final positions. Since the SiPM gain is sensitive to temperature, a cooling system was used to remove the heat generated by the electronics. The cooling system consisted of multiple water coils connected to cold plates. The plates were coupled to the preamplifier boards that follow the SiPMs. Both the cooling system and electronics were controlled remotely. The EMCal prototype can be seen in Figure \ref{setup_emcal}, which shows the blocks, light guides, SiPMs, electronics and part of the cooling system.

\section{Readout Electronics and Data Acquisition}
\label{sec:electronics}

The summed signals from the four SiPMs from a tower were sent to a preamplifier, then shaped and driven into a digitizer. The SiPMs were operated at 4V above their breakdown voltage, which produces a gain of approximately $2.3 \times 10^5$. A small thermistor was mounted at the center of the four SiPMs to monitor the temperature per tower. The temperature of the SiPMs was held constant within approximately 0.5$^{\circ}$C. Since the gain temperature dependence of the SiPMs is approximately 1.5$\%$/$^{\circ}$C, temperature variations did not contribute significantly to the measured energy resolution. LEDs with an emission peak at 405 nm were mounted near the readout end of each tower and were used to provide a pulsed light source for calibration. Similarly, a charge injection test pulse was used to test and calibrate the readout electronics. The EMCal prototype could operate in a nominal gain mode, or a high gain mode with 16 times the normal gain. The gain was selected through a slow control system.

The slow control system consisted of an interface board connected to a controller board. The interface board was mounted on the EMCal prototype while the controller board was in a separate crate. The interface board contained digital-to-analog converters needed for different testing and monitoring tasks. The interface board controlled the SiPM bias and gain. Testing of the preamplifiers was controlled through the interface board as well. The interface board also monitored leakage current and local temperature for compensation. The parameters for these testing and monitoring tasks were provided to the interface board by the controller board. An ethernet connection was used to communicate with the controller board.

Signals were digitized using a digitization system developed for PHENIX \cite{Anderson:2011}. The signal waveforms were digitized using Analog-to-digital converters (ADC) at a sampling frequency of 60 MHz, followed by Field Programmable Gate Arrays. Signals were collected in Data Collection Modules and the data was finally recorded using the data acquisition system RCDAQ \cite{RCDAQ2,RCDAQ}. The signals were recorded for the EMCal prototype as well as the external detectors mentioned in section \ref{sec:testbeam}.

\section{Test Beam}
\label{sec:testbeam}

The EMCal prototype was tested at the Fermilab Test Beam Facility as experiment T-1044. The facility provided a particle beam, detectors such as a lead-glass calorimeter and Cherenkov counters, and a motion table in the MT6.2C area \cite{ftbf}. The EMCal was placed on the motion table to allow testing in different positions with respect to the beam.

The particle beam used in the experiment had energies ranging from 2 to 28 GeV and a profile size of a few centimeters, dependent on beam energy. The beam was composed mainly of electrons, muons, and pions, and their relative abundance depended on the energy \cite{Feege:2011dsa,Blatnik:2015bka}. The beam hit the EMCal prototype with a frequency of 1 spill per min, where a spill corresponds to a maximum of approximately $10^5$ particles during 4 seconds. The beam had a nominal momentum spread of $\delta p/p \approx 2\%$ for the energy range used \cite{Tsai:2012cpa,Aidala:2017rvg,meson_mom}. A lead-glass calorimeter was used to measure the average and the spread of the beam momentum. The lead-glass calorimeter had a size of $45 \times 15 \times 15$~cm$^3$ and an approximate resolution of $1.4 \% \oplus 5.0\%/\sqrt{E}$~\cite{Aidala:2017rvg}.

External detectors were used to discriminate electron signals from minimum ionizing particles (MIPs) and hadrons. Two gaseous Cherenkov counters were used for particle identification. The gas pressure in each Cherenkov counter was tuned to trigger only on electron signals. A hodoscope \cite{Tsai:2012cpa,Tsai:2015bna} was placed upstream of the EMCal to determine the position of the particles in the beam precisely. The hodoscope consisted of 16 hodoscope fingers (0.5 cm wide scintillators) arranged in two arrays of 8 fingers each. One array had the hodoscope fingers arranged vertically and the other array had them arranged horizontally. The position of a hit in the hodoscope was given by a horizontal and a vertical hodoscope channel number. Each hodoscope finger was read out by an SiPM. Four veto detectors were also placed around the EMCal in order to suppress particles traveling outside the acceptance of the hodoscope. Each veto counter consisted of a scintillator coupled to a photomultiplier tube (PMT).

\section{Simulations}
\label{sec:simulations}

The EMCal prototype was simulated using \mbox{\sc Geant4}\xspace \cite{Agostinelli:2002hh,Allison:2006ve} version 4.10.02-patch-02, with the physics list \mbox{\sc QGSP\_BERT}. The EMCal blocks were simulated following their nominal design with a uniform block density. The simulations included an electron beam with a Gaussian profile. An 8 GeV beam with a standard deviation of 8 cm was used to study the prototype's energy response as a function of position. To study the prototype's energy response as a function of energy, the beam had an energy between 2 and 28 GeV and a standard deviation of 2.5 cm. For this energy dependent study, the beam was pointed between Towers A and B, which are located near the center of the prototype (see Figure \ref{towers}). In the simulations, the energy deposits from the electromagnetic showers were converted into light using Birks' law \cite{Birks:1951boa} with constant $k_B = $ 0.0794 mm/MeV \cite{Hirschberg:1992xd}. The number of photons collected was reduced by the light guide collection efficiency and then converted to number of fired SiPM pixels taking into account the SiPM saturation. The saturation was simulated by considering a Poisson distribution of photons randomly hitting the pixels and counting the total number of fired pixels. The mean of the Poisson distribution was proportional to the beam input energy, giving an energy dependent saturation effect. The number of fired pixels was converted to ADC counts and then calibrated to an input energy. The simulations were integrated into the sPHENIX analysis framework.

\section{Analysis Methods}
\label{sec:analysis}

\subsection{Data Sets}

\begin{figure}[htb!]
  \centering
  \includegraphics[width=0.4\textwidth]{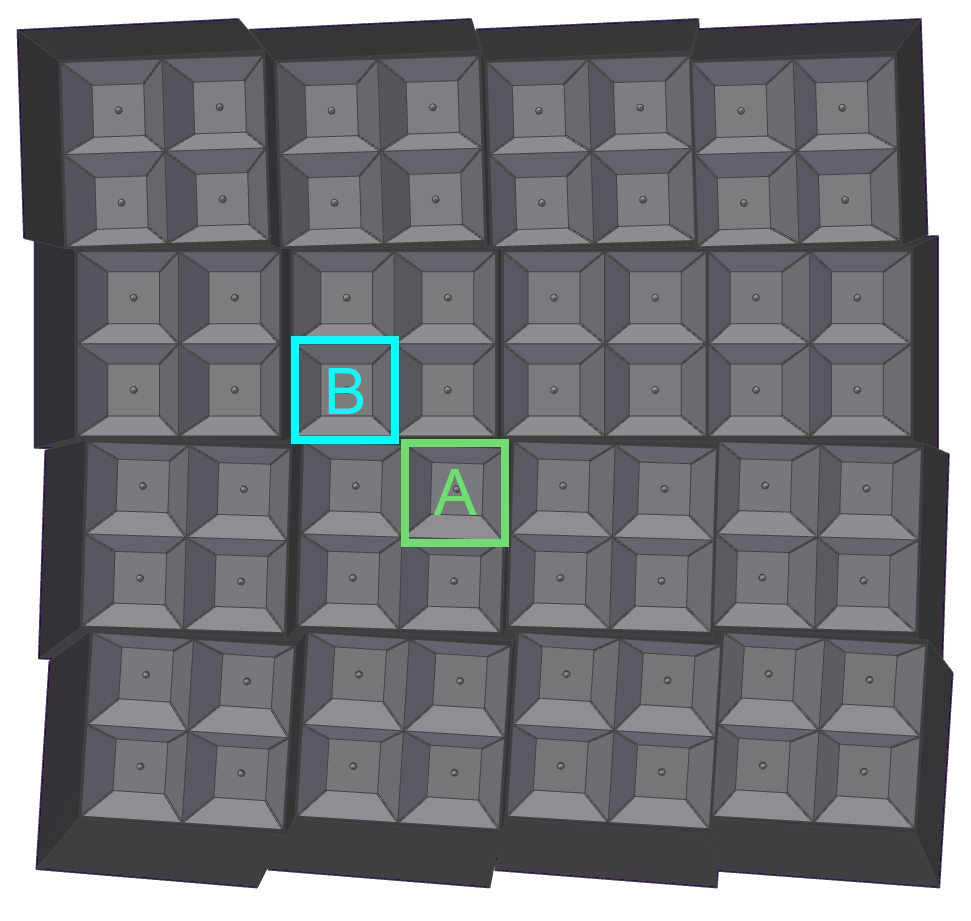}
  \caption{Front view of the EMCal prototype showing the towers. Tower A (light green) and Tower B (light blue) are highlighted.}
  \label{towers}
\end{figure}

The data sets used in this analysis correspond to a beam of electrons with energies of 2, 3, 4, 6, 8, 12, 16, 20, 24 and 28 GeV. The beam was pointed at either Tower A or Tower B (see Figure \ref{towers}). In this paper, whenever Tower A or Tower B is mentioned, it is referring to the corresponding data set that had the beam centered at either of those towers.

\subsection{Electron Selection}
Various cuts were used in order to suppress MIPs and hadrons, and select only events with single electrons. Single electrons were identified by requiring a Cherenkov cut, a vertical and horizontal hodoscope cut, and four veto cuts. It was generally assumed that the high energy peak in the energy spectra of the Cherenkov counters and hodoscope channels corresponded to the electrons. For the veto cuts, the high energy peak was assumed to correspond to particles traveling outside the beam position. The Cherenkov cut required the pulse height in the Cherenkov counters to be consistent with that of an electron. For the vertical and horizontal hodoscope cuts, the events were required to have an energy greater than 50$\%$ of the peak energy in each hodoscope finger's energy spectrum. Only events with one hit in the vertical and one hit in the horizontal hodoscope fingers were considered. For the four veto cuts, the events were required to have an energy less than 20$\%$ of the peak energy in each veto detector's energy spectrum. These cuts gave a number of single electrons of approximately 5,000-50,000, depending on the energy.

\subsection{Calibration}
\begin{figure*}[hbt!]
  \centering
  \includegraphics[width=\textwidth]{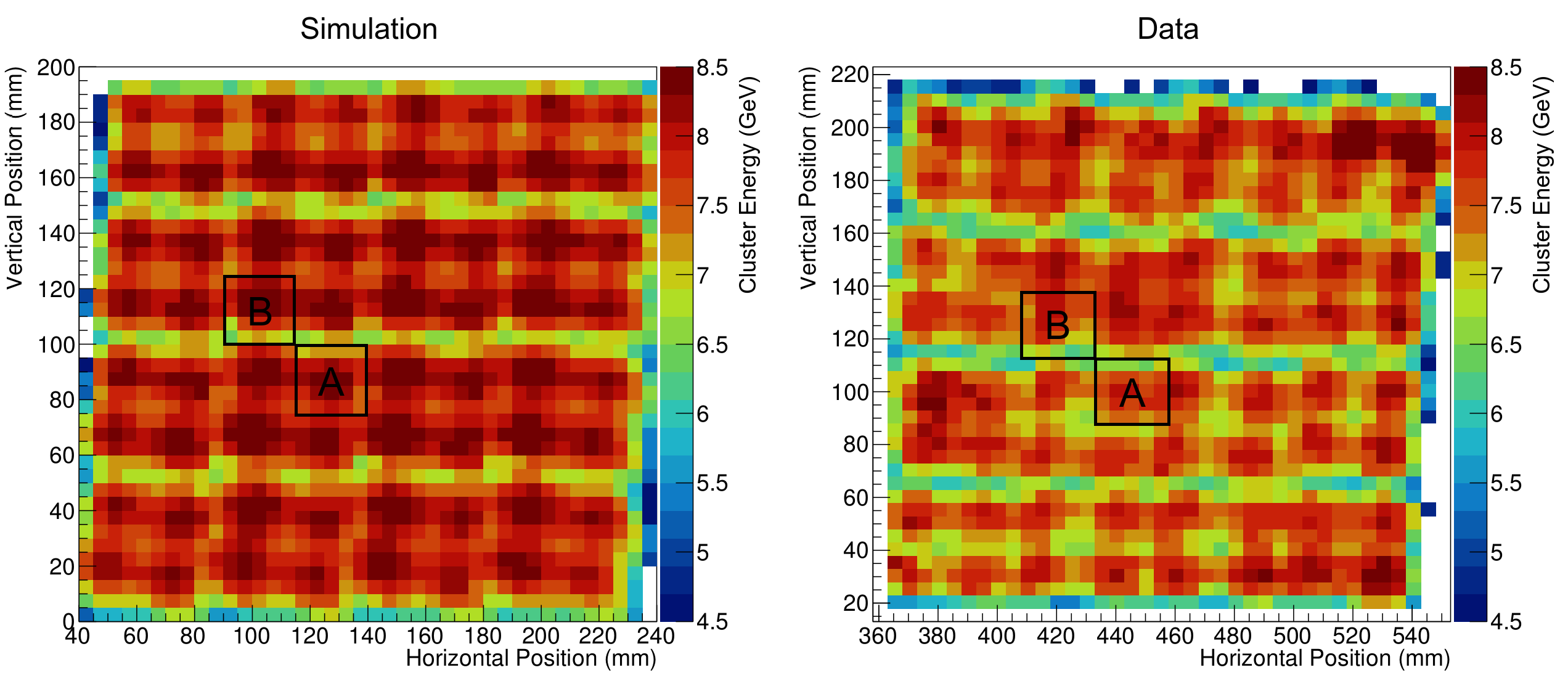}
  \caption{Cluster Energy vs. Position for simulations (left panel) and data (right panel). The results correspond to an input energy of 8 GeV. Towers A and B are shown in black squares.}
  \label{uniformity}
\end{figure*}

A preliminary calibration of the data, termed the \emph{shower calibration}, was performed based on how the electromagnetic showers develop within the EMCal. A uniformity study of the EMCal prototype showed that the energy measurements depend on the transverse position within the EMCal. Figure \ref{uniformity} shows the measured energy as a function of position for an input energy of 8 GeV, for both data and simulations. A higher energy response was observed towards the center of the towers than at the boundaries between the blocks and towers. This behavior motivated the use of secondary energy calibrations, the \emph{position dependent correction} and the \emph{beam profile correction}.

The calibration procedures are as follows:

\subsubsection{Shower calibration}

For each event, the energy measured by the EMCal was obtained as the total energy of a 5$\times$5 cluster of towers around the maximum energy tower. The size of the cluster was selected based on the Moli\`ere radius for the EMCal blocks. A cluster of 5$\times$5 towers contains over 95$\%$ of the electromagnetic shower energy. The energy corresponding to a cluster of 5$\times$5 towers around the tower with the maximum energy is called the \emph{cluster energy} and is denoted as $E_{\rm cluster}$. The average cluster energy for an 8 GeV electron beam incident at the center of each tower was reconstructed to the input energy and calibration constants were applied tower-by-tower.

\subsubsection{Position dependent correction}

The energy measured by the EMCal was corrected by a constant that depends on the position of the hit in the EMCal. Two different corrections were obtained, the difference lying in the availability of external position information. In the first, the position was determined by a horizontal and a vertical hodoscope finger, with a total of 8$\times$8 possible positions. In the second, the position was determined by the energy averaged cluster position measured by the EMCal, discretized in 8$\times$8 bins that matched the hodoscope. The position dependent calibration constants were obtained from 8 GeV data as described below. The procedure is the same for both the hodoscope-based and cluster-based corrections. For each of the 64 possible position bins, a histogram was filled with the cluster energy in that position. The histogram was then fit with a Gaussian of mean $\mu$. The calibration constant for each position was obtained as 8 GeV/$\mu$. The position dependent correction improved the energy resolution by 2-3$\%$, depending on the energy.

\begin{figure*}[htb!]
  \centering
  \includegraphics[width=\textwidth]{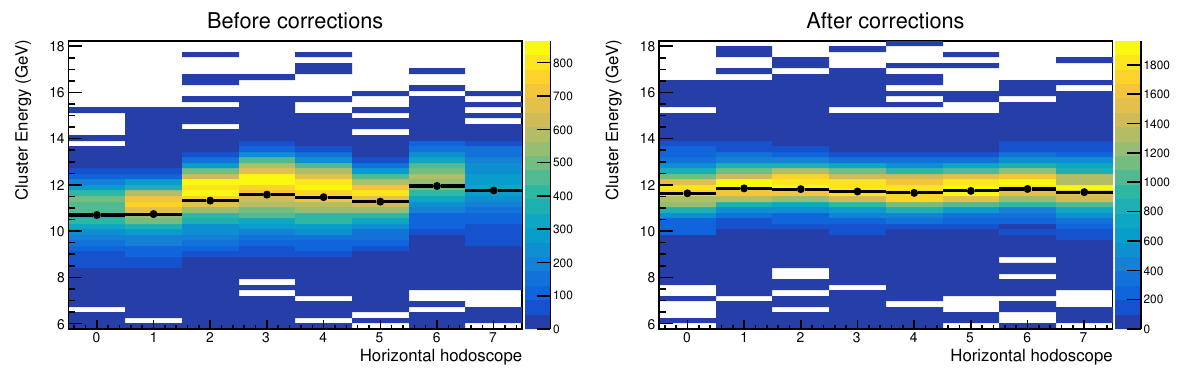}
  \caption{Cluster Energy vs. Horizontal Hodoscope Position before (left panel) and after (right panel) applying the hodoscope-based position dependent correction and the beam profile correction. The color scale represents the number of events, while the black points correspond to the mean of the energy distributions for each hodoscope position. The data corresponds to a 12 GeV beam centered at Tower A.}
  \label{before_after_hodo_correction}
\end{figure*}

The sPHENIX tracker can be used in place of a hodoscope to develop a position dependent correction. Since the tracker is only sensitive to charged particles, the cluster-based correction can be used for neutral particles instead.

\subsubsection{Beam profile correction}

In the experiment, the beam had a different transverse profile at different energies. In addition to the position dependent correction, a \textit{beam profile correction} was introduced in order to correct for the energy dependence of the beam profile. This correction consisted of filling the energy histograms with weights that were obtained by making the distribution of beam particles uniform as a function of position. The beam profile correction changed the energy resolution by 0.1-0.5$\%$, depending on the energy.

The effects of these corrections on the energy response can be seen in Figure \ref{before_after_hodo_correction}. This figure shows the cluster energy as a function of horizontal hodoscope position. The data is shown before and after applying the hodoscope-based position dependent correction and the beam profile correction. After the corrections are applied, the energy response of the EMCal becomes more uniform.

The simulations also included the position dependent and beam profile corrections. The corrections were obtained using the procedure previously described, where the simulated position was discretized in 8$\times$8 bins to mock the hodoscope.

\section{Results and Discussion}
\label{sec:results}
 
Following the analysis procedure described in the previous section, the energy resolution and linearity of the EMCal prototype was obtained for input energies ranging from 2 to 28~GeV, for both simulations and data.

\begin{figure*}[ht!]
\centering
\includegraphics[width=\textwidth]{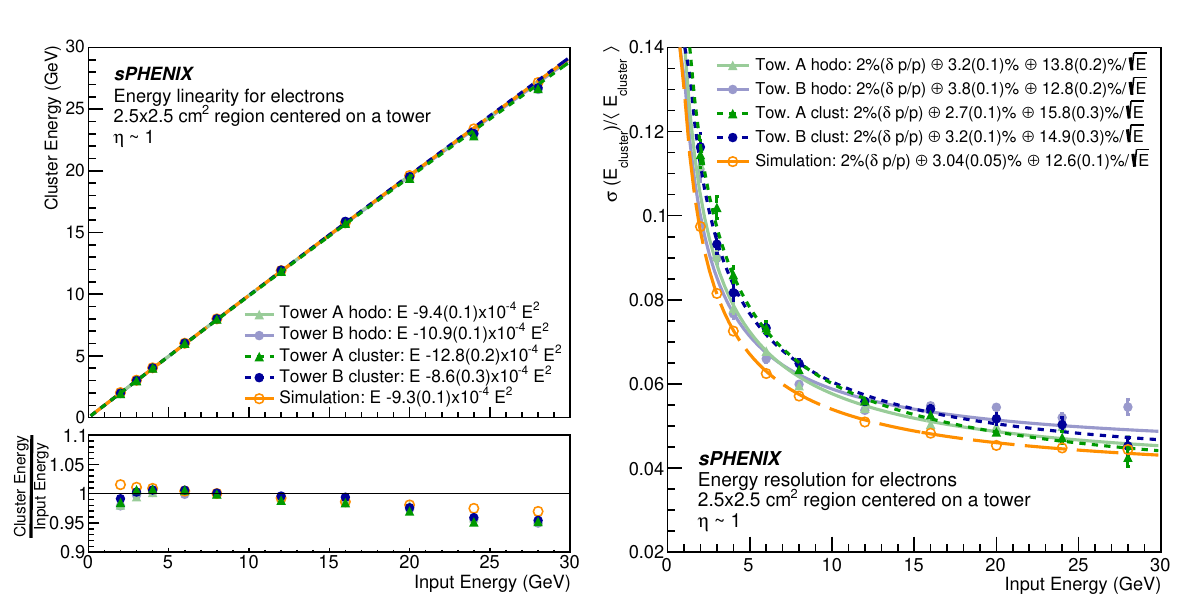}
\caption{Linearity and resolution of the EMCal prototype for a $2.5\times2.5$ cm$^2$ cut centered on a tower. The $2.5\times2.5$~cm$^2$ cut was selected based on the approximate area of a tower. The data corresponds to Tower A (green triangles) and Tower B (purple full circles). The data was corrected using the hodoscope-based (solid lines) and cluster-based (fine dashed lines) position dependent corrections, as well as the beam profile correction. Simulations (orange open circles, coarse dashed line) are shown for comparison and include the same corrections as the data. (top left panel) Cluster Energy vs. Input Energy. (bottom left panel) $\frac{\text{Cluster Energy}}{\text{Input Energy}}$ vs. Input Energy. The linearity was obtained as $E_{\rm cluster} = E+c E^2$. (right panel) Energy Resolution vs. Input Energy. The resolution was obtained as $\sigma(E_{\rm cluster})/\langle E_{\rm cluster}\rangle = \delta p/ p \oplus a \oplus b/\sqrt{E}$, where a $\delta p/p = 2 \%$ term was added to account for the beam momentum spread. } \label{resolution_2es_w_clusterhodosim_hodcut5}
\end{figure*}

Figure \ref{resolution_2es_w_clusterhodosim_hodcut5} shows the energy resolution and linearity of the EMCal prototype using a $2.5\times2.5$~cm$^2$ cut centered at the tower. The $2.5\times2.5$~cm$^2$ cut was selected based on the approximate area of a tower. The results are shown for data and simulations and include all corrections. The uncertainty bars on the data points correspond to the statistical uncertainties. The linearity was obtained as $E_{\rm cluster} = E+c E^2$, where $E$ is the input energy and $c$ is a constant. The resolution was obtained as $\sigma(E_{\rm cluster})/\langle E_{\rm cluster}\rangle = \delta p/ p \oplus a \oplus b/\sqrt{E}$, where $a$ and $b$ are constants, and a $\delta p/p = 2 \%$ term was added to account for the beam momentum spread. Table \ref{fit_constants_5x5} shows the values of the fit constants $a, b$, and $c$.

\begin{table}[!hbt]
\caption{EMCal Energy Linearity and Resolution for a $2.5\times2.5$ cm$^2$ cut centered on a tower}
Resolution fit: $\sigma(E_{\rm cluster})/\langle E_{\rm cluster}\rangle = 2 \% \oplus a \oplus b/\sqrt{E}$ \\
\\
Linearity fit: $E_{\rm cluster} = E+c E^2$\\
\\
\begin{tabularx}{.5\textwidth}{l@{}c@{}c@{}c@{}c}
\hline
\multicolumn{1}{c}{\bf } & \multicolumn{1}{c}{Tower} & \multicolumn{1}{c}{\bf $a$ ($\%$)} & \multicolumn{1}{c}{ $b$ ($\%$ GeV$^{1/2}$)} & \multicolumn{1}{c}{ $c$ (GeV$^{-1}$)} \\
\hline
\\
Data, hodoscope & A & 3.2 $\pm$ 0.1 & 13.8 $\pm$ 0.2 & (-9.4 $\pm$ 0.1)$\times 10^{-4}$\\
\\
Data, hodoscope & B & 3.8 $\pm$ 0.1 & 12.8 $\pm$ 0.2 & (-10.9 $\pm$ 0.1)$\times 10^{-4}$\\
\\
Data, cluster & A & 2.7 $\pm$ 0.1 & 15.8 $\pm$ 0.3 & (-12.8 $\pm$ 0.2)$\times 10^{-4}$\\
\\
Data, cluster & B & 3.2 $\pm$ 0.1 & 14.9 $\pm$ 0.3 & (-8.6 $\pm$ 0.3)$\times 10^{-4}$\\
\\
Simulation & & 3.04 $\pm$ 0.05 & 12.6 $\pm$ 0.1 & (-9.3 $\pm$ 0.1)$\times 10^{-4}$\\
\\
\hline
\end{tabularx}
\label{fit_constants_5x5}
\end{table}

\begin{figure*}[ht!]
\centering
\includegraphics[width=\textwidth]{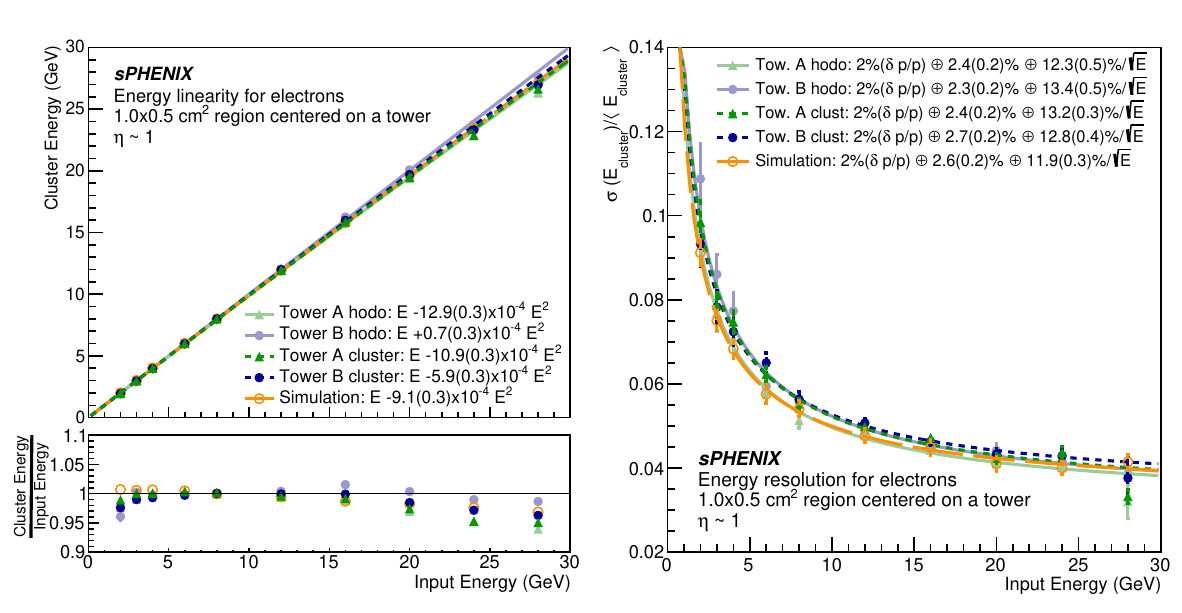}
\caption{Linearity and resolution of the EMCal prototype for a $1.0\times0.5$ cm$^2$ cut at the center of a tower. The data corresponds to Tower A (green triangles) and Tower B (purple full circles). The data was corrected using the hodoscope-based (solid lines) and cluster-based (fine dashed lines) position dependent corrections, as well as the beam profile correction. Simulations (orange open circles, coarse dashed line) are shown for comparison and include the same corrections as the data. (top left panel) Cluster Energy vs. Input Energy. (bottom left panel) $\frac{\text{Cluster Energy}}{\text{Input Energy}}$ vs. Input Energy. The linearity was obtained as $E_{\rm cluster} = E+c E^2$. (right panel) Energy Resolution vs. Input Energy. The resolution was obtained as $\sigma(E_{\rm cluster})/\langle E_{\rm cluster}\rangle = \delta p/ p \oplus a \oplus b/\sqrt{E}$, where a $\delta p/p = 2 \%$ term was added to account for the beam momentum spread.} \label{resolution_2es_w_clusterhodosim_hodcut21}
\end{figure*}

The resolution obtained with the cluster-based correction differs from the hodoscope-based correction by approximately 0.6$\%$ in the constant term and 2.1$\%$ in the $1/\sqrt{E}$ term. Since the cluster-based correction depends on the position measured by the EMCal itself and not the hodoscope, the difference in the results can potentially arise from the reduced cluster position resolution of the EMCal at lower energy. Additionally, the energy resolution seems to be better in the simulations than in the hodoscope corrected data by approximately 0.5$\%$ in the constant term and 0.7$\%$ in the $1/\sqrt{E}$ term. These differences can arise from the lower energy collection efficiency at the boundaries between towers and blocks, as well as tower by tower variations that are not present in the simulations. The differences in the resolution results can be minimized by making a cut at the center of the towers, where the energy collection is most efficient. Figure \ref{resolution_2es_w_clusterhodosim_hodcut21} shows the linearity and resolution results using a 1.0$\times$0.5 cm$^2$ cut at the center of the towers. This figure shows better agreement between data and simulations. Table \ref{fit_constants_2x1} shows the corresponding linearity and resolution fit constants.

\begin{table}[!hbt]
\caption{EMCal Energy Linearity and Resolution for a $1.0\times0.5$ cm$^2$ cut at the center of a tower}
Resolution fit: $\sigma(E_{\rm cluster})/\langle E_{\rm cluster}\rangle = 2 \% \oplus a \oplus b/\sqrt{E}$ \\
\\
Linearity fit: $E_{\rm cluster} = E+c E^2$\\
\\
\begin{tabularx}{.5\textwidth}{l@{}c@{}c@{}c@{}c}
\hline
\multicolumn{1}{c}{\bf } & \multicolumn{1}{c}{Tower} & \multicolumn{1}{c}{\bf $a$ ($\%$)} & \multicolumn{1}{c}{ $b$ ($\%$ GeV$^{1/2}$)} & \multicolumn{1}{c}{ $c$ (GeV$^{-1}$)} \\
\hline
\\
Data, hodoscope & A & 2.4 $\pm$ 0.2 & 12.3 $\pm$ 0.5 & (-12.9 $\pm$ 0.3)$\times 10^{-4}$\\
\\
Data, hodoscope & B & 2.3 $\pm$ 0.2 & 13.4 $\pm$ 0.5 & (+0.7 $\pm$ 0.3)$\times 10^{-4}$\\
\\
Data, cluster & A & 2.4 $\pm$ 0.2 & 13.2 $\pm$ 0.5 & (-10.9 $\pm$ 0.3)$\times 10^{-4}$\\
\\
Data, cluster & B & 2.7 $\pm$ 0.2 & 12.8 $\pm$ 0.4 & (-5.9 $\pm$ 0.3)$\times 10^{-4}$\\
\\
Simulation & & 2.6 $\pm$ 0.2 & 11.9 $\pm$ 0.3 & (-9.1 $\pm$ 0.3)$\times 10^{-4}$\\
\hline
\end{tabularx}\label{fit_constants_2x1}
\end{table}

Additionally, Figure \ref{resolution_2es_w_clusterhodosim_hodcut5} shows that for energies below 15 GeV the energy resolution for Towers A and B generally agree within the statistical uncertainties, while for higher energies the resolution is consistently larger for Tower B than for Tower A. The disagreement between the resolution of the towers above 15 GeV is observed for both the hodoscope-based and cluster-based results of Figure \ref{resolution_2es_w_clusterhodosim_hodcut5} and contributes to the fit constants of Table \ref{fit_constants_5x5}. However, this disagreement is not observed when a cut at the center of the towers is used, as shown in Figure \ref{resolution_2es_w_clusterhodosim_hodcut21} and Table \ref{fit_constants_2x1}.

Comparing the 2018 results to the 2016 results of reference \cite{Aidala:2017rvg}, the resolution improved for energies in the range 2 to 8 GeV. In terms of the resolution fit, the $1/\sqrt{E}$ term of the resolution decreased by approximately 2.5$\%$ and the constant term increased by approximately 0.7$\%$. Furthermore, the linearity improved by approximately 1$\%$ in the 2018 prototype with respect to the 2016 prototype.

\section{Conclusions}
\label{sec:conclusions}

A 2D projective prototype of the sPHENIX EMCal was constructed and tested. The EMCal prototype's energy response to electrons was studied as a function of incident position and energy. The energy resolution and linearity of the EMCal prototype were obtained using two different position dependent energy corrections (hodoscope-based and cluster-based) as well as a beam profile correction. The two data sets used in this analysis had beam energies ranging from 2 to 28 GeV, but one had the beam centered at Tower A and the other one had the beam centered at Tower B. The energy resolution was obtained for each tower using a cut of $2.5\times2.5$ cm$^2$ centered on the tower. Based on the hodoscope position dependent correction, the EMCal prototype was found to have a tower averaged energy resolution of $\sigma(E)/\langle E\rangle = 3.5(0.1) \oplus 13.3(0.2)/\sqrt{E}$. Based on the cluster position dependent correction, the tower averaged energy resolution was found to be $\sigma(E)/\langle E\rangle = 3.0(0.1) \oplus 15.4(0.3)/\sqrt{E}$. These energy resolution results meet the requirements of the sPHENIX physics program.

\section*{Acknowledgments}

The authors would like to thank the technical staffs of the University of Illinois at Urbana-Champaign and Brookhaven National Laboratory for assisting the construction of the electromagnetic calorimeter prototype. The authors would also like to thank Dr. O. Tsai for providing the hodoscope used in the beam test.

\thanks{C.A.~Aidala, E.A.~Gamez, N.A.~Lewis and J.D.~Osborn are with the Department of Physics at the University of Michigan, Ann Arbor, MI 48109-1040.}

\thanks{S.~Altaf, M.~Phipps, C.~Riedl, T.~Rinn, A.C.~Romero~Hernandez, A.M.~Sickles, E.~Thorsland and X.~Wang are with the Department of Physics at the University of Illinois Urbana-Champaign, Urbana, IL 61801-3003.}

\thanks{R.~Belmont is with the Department of Physics at the University of Colorado Boulder, Boulder, CO 80309-0390 and the Department of Physics and Astronomy at the University of North Carolina Greensboro, Greensboro, NC 27402-6170.}

\thanks{S.~Boose, D.~Cacace, E.~Desmond, J.S.~Haggerty, J.~Huang, M.D~Lenz, W.~Lenz, E.J.~Mannel, R.~Pisani, S.~Polizzo, M.~Purschke, S.~Stoll, and C.L.~Woody are with Brookhaven National Laboratory, Upton, NY 11973-5000.}

\thanks{M. Connors is with the Department of Physics and Astronomy at the Georgia State University, Atlanta, GA 30302-5060 and the RIKEN BNL Research Center, Upton, NY 11973-5000.}

\thanks{J.~Frantz and A.~Pun are with the Department of Physics and Astronomy at Ohio University, Athens, OH 45701-6817.}

\thanks{N.~Grau is with the Department of Physics at Augustana University, Sioux Falls, SD 57197.}

\thanks{A.~Hodges, M. Sarsour and X. Sun are with the Department of Physics and Astronomy at the Georgia State University, Atlanta, GA 30302-5060.}

\thanks{Y.~Kim is with the Department of Physics at the University of Illinois Urbana-Champaign, Urbana, IL 61801-3003 and the Department of Physics and Astronomy at Sejong University, Seoul 05006, Korea.}

\thanks{D.V.~Perepelitsa, C.~Smith, and F.~Vassalli are with the Department of Physics at the University of Colorado Boulder, Boulder, CO 80309-0390.}

\thanks{Z.~Shi is with the Department of Physics at the Massachusetts Institute of Technology, Cambridge, MA 02139-4307.}

\bibliographystyle{IEEEtran}

\bibliography{IEEEabrv,references}

\end{document}